# Origin of Asteroid (101955) Bennu and its Connection to the New Polana Family

Driss Takir*, Jacobs, NASA Johnson Space Center, Houston, TX
Joshua P. Emery, Northern Arizona University, Flagstaff, AZ
William F. Bottke, Southwest Research Institute, Boulder, CO
Anicia Arredondo, Southwest Research Institute, San Antonio, TX

*Corresponding Author: driss.takir@nasa.gov

## Abstract

The asteroid (142) Polana is classified as a B-type asteroid located in the inner Main Belt. This asteroid is the parent of the New Polana family [1], which has been proposed to be the likely source of primitive near-Earth asteroids such as the B-type asteroid (101955) Bennu [2, 3]. To investigate the compositional correlation between Polana and Bennu at the 3-μm band and their aqueous alteration histories, we analyzed the spectra of Polana in the ~2.0-4.0-μm spectral range using the NASA Infrared Telescope Facility in Hawai'i. Our findings indicate that Polana does not exhibit discernable 3-μm hydrated mineral absorption (within 2σ), which is in contrast to asteroid Bennu. Bennu displayed a significant 3-μm absorption feature similar to CM- and CI-type carbonaceous chondrites [4]. This suggests two possibilities: either Bennu did not originate from the New Polana family parented by asteroid Polana or the interior of Bennu's parent body was not homogenous, with diverse levels of aqueous alteration. Several explanations support the latter possibility, including heating due to shock waves and pressure, which could have caused the current dehydrated state of Bennu's parent body.

## Introduction

Near-Earth Asteroids (NEAs) are minor bodies with diameters ranging from meteorite-sized objects to bodies that are tens of kilometers in diameter. They represent almost all asteroid taxonomic classes in the Inner Main Belt (IMB) region, defined with semimajor axes between 2.1 and 2.5 au [5]. NEAs' average collisional lifetime is $\leq 10^7$ years, much shorter than the age of the solar system [3]. They reached the mean-motion and secular resonances in the IMB by Yarkovsky drift or by collisions and were delivered to their current near-Earth orbits via gravitational perturbations within these resonances [6, 7]. About 60 to 70% of sub-kilometer NEAs and about 30% of the multikilometer NEAs come from the IMB, according to numerical models of the NEA population [8]. Seven primitive collisional families have been identified in the low-inclination region of the IMB: New Polana, Eulalia, Erigone, Sulamitis, Clarissa, Chaldaea, and Klio that might be the source of primitive NEAs [8, 3]. This study aims to investigate the compositional connection between asteroid Polana and Bennu, to determine if Bennu originated from the New Polana family parented by asteroid Polana or if the interior of the original parent body was heterogeneous.

Collisional events, including disruptive impacts between asteroid-sized objects, have dominated our solar system's history and played a significant role in forming asteroid families. The New Polana family is a low-inclination and the most prominent low-albedo family within the IMB between the $v_6$ secular resonance, which marks the IMB's innermost boundary (near 2.1 to 2.2 au for low inclinations) and the 3:1 mean-motion resonance with Jupiter at ~2.5 AU [1]. This family is parented by the B-type asteroid (142) Polana [1] and has an age of $1400 \pm 150$ Myr [3]. Primitive near-Earth asteroids (NEAs), including Hayabusa2's asteroid target (162173) Ryugu and OSIRIS-REx's asteroid target (101955) Bennu, were likely formed from collisional disrupted fragments during the New Polana family formation event [2, 3]. The background asteroid population (i.e., asteroids that do not cluster to form families), which includes about two times more km-sized objects than the New Polana family, was also suggested to be the delivery source of more primitive objects to the $v_6$ secular resonance than any low-albedo family [2].

Several observers studied the largest primitive IMB families in the Visible (VIS: ~0.3-1.0 µm) [9, 10, 11,12, 13] and found that these primitive families could be classified into two main spectral groups: The Erigone-like group is spectrally diverse, dominated by C-type asteroids, and exhibits a hydration feature at 0.7 µm, a charge transfer feature attributed to oxidized iron in phyllosilicates [14]. The New Polana- and Eulalia-like group is spectrally homogeneous, contrary to what [1] predicted, and shows a broad 1-µm feature instead of the hydration feature at 0.7 µm. Additionally, using near-infrared (NIR: ~0.8-2.5 µm) spectra, [15, 16, 17] found no distinct diagnostic absorption features that indicate spectral surface composition diversity in the Sulamitis, Chaldaea, and Klio families.

Asteroid 142 Polana, the largest remnant of the New Polana family, has been spectrally (~0.5-2.5 µm) and dynamically linked to asteroid Bennu [2, 3, 9, 18]. Bennu's spectra were measured over the wavelength range from 0.4 to 4.3 µm with OSIRIS-REx's Visible and InfraRed Spectrometer (OVIRS) [19]. Spectra of Bennu are similar to those of aqueously altered CM-type carbonaceous chondrites in the 3-µm band [4]. Here, we investigate the compositional linkage of asteroids Polana and Bennu in the 3-µm band using the Infrared Telescope Facility (IRTF) in Hawai'i. If Polana shows a 3-µm band consistent with the one detected on Bennu, it would support the hypothesis that the two asteroids are compositionally related and experienced the same aqueous alteration environments. Otherwise, justifications must be provided to explain the discrepancy between Polana and Bennu in the 3-µm band and why we consider that the New Polana family is the most probable source for Bennu and other primitive NEAs.

**Methods**

*Observational Technique*

The present study includes spectra of Polana that were measured using the prism (0.7- 2.5 µm) and the long-wavelength cross-dispersed (LXD: 1.9– 4.2-µm) modes of the SpeX spectrograph/imager at IRTF [20] (Table 1). This investigation also includes previously published visible spectra (0.4–0.93-µm) of Polana [21, 22] to help provide a deeper understanding of its surface composition. The Interactive Data Language (IDL)-based spectral extraction and reduction tool Spextool version v 4.1 (https://irtfweb.ifa.hawaii.edu/~spex/observer/) was used to reduce Polana's spectral data [23].

Spectra of G-dwarf stars (HD 216801 and HD 98737) with solar-like B-V and V-K colors were used as standard stars in both the prism and LXD modes to correct for the telluric water vapor absorption features in Polana's spectra (Table 1). The spectral image frames of Polana and standard stars at beam position A were subtracted from the spectral image frames at beam B of the telescope to remove the background sky. Wavelength calibration was conducted using argon lines, measured with internal calibration for the prism and LXD modes using telluric absorption lines. The flat field frames were generated by illuminating an integrating sphere in the calibration box.

**Table 1.** Observing parameters for asteroid (142) Polana observed with prism and LXD modes of SpeX at NASA IRTF.

| SpeX mode | SpeX slit (arcsec) | Date (UT) | Time (UT) | Airmass | Standard star | Spectral type | B-V* | V-K* |
|---|---|---|---|---|---|---|---|---|
| **Prism** | 1.6 x 15 | 07/02/2023 | 12:32-12:57 | 1.30-1.22 | HD 216801 | G0V | 0.59 | 1.55 |
| **LXD** | 0.8 x 15 | 03/02/2022 | 8:02-1:30 | 1.57-1.07 | HD 98737 | G5 | 0.62 | 1.35 |

* B–V and V–K represent stars' colors.

*Thermal Modeling and Correction*

To model the thermal flux longwards of 2.5 μm in Polana's spectra, we used the Near-Earth Asteroid Thermal Model (NEATM), which was developed by [24] based on the Standard Thermal Model (STM) [25]. We fitted the measured thermal excess of Polana with a model excess (Figure 1, left), which was subtracted from Polana's measured thermal flux relative to LXD spectra (Figure 1, right). Parameters, including the heliocentric and geocentric distances, visible geometric albedo, and phase angle at Polana's observation, were acquired from the Jet Propulsion Laboratory (JPL) Horizon online ephemeris generator (Table 2, https://ssd.jpl.nasa.gov/horizons/). We used a default value for Polana's slope parameter G of 0.15 [26]. The NEATM model includes the beaming parameter ($\eta$) to accommodate variations in Polana's surface roughness. The initial value for the beaming parameter was set to 1. However, the NEATM model went through multiple iterations of beaming parameter and geometric albedo approximations to minimize the chi-squared fit to observational data of Polana and determine the surface temperature that most closely aligns with the measured thermal flux. Both bolometric and spectral emissivity were assumed to be 0.9, and the night side of Polana emits no thermal energy in the model.

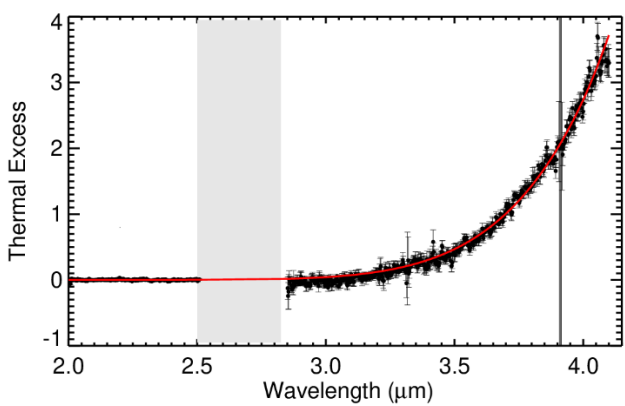
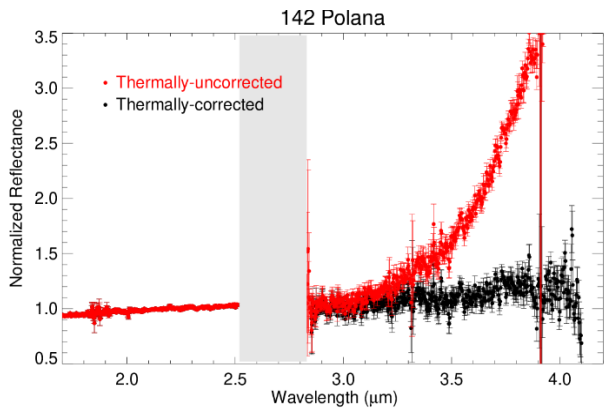

**Figure. 1. (Left)** Thermal excess and the best-fit NEATM thermal model (in red) for asteroid Polana. Uncertainties were computed by Spextool software using the Robust Weighted Mean algorithm with a clipping threshold of 8 (sigma). The value at each pixel is the weighted average of the good pixels, and the propagated variance gives the uncertainty. **(Right)** Thermally uncorrected (red) and corrected (black) spectrum of asteroid Polana. The gray bars on each plot indicate wavelengths of strong absorption by water vapor in Earth's atmosphere.

**Table 2.** Asteroid thermal model inputs are used to model and correct the thermal excess in (142) Polana.

| Rau[*] (AU) | Dau[*] (AU) | Phase angle (deg) | Geometric albedo | H mag | Rotation period (h) | K/V [**] | Temp[***] (K) |
|---|---|---|---|---|---|---|---|
| 2.21 | 1.23 | 4.52 | 0.047 | 7.23 | 10.60 | 1.04 | 271 |

\* Rau is the heliocentric range, and Dau is the geocentric range.

\*\* K-band to V-band scale, applied to spectra to reconcile the two reflectance values at two different wavelengths.

\*\*\* Sub-solar temperature corresponds to the optimum thermal model.

*Spectral Analysis*

We used a standard technique in the field for spectral analysis (e.g., [27]). We calculated the band depth $D_{2.90}$, at a wavelength of 2.90 μm, relative to the continuum (the linear regression across the K-band: 1.90-2.50 μm):

$$D_{2.90} = (R_c - R_{2.90}) / R_c \qquad (1)$$

where $R_{2.90}$ is the reflectance at a given wavelength of 2.90 μm, and $R_c$ corresponds to the reflectance of the continuum at the same wavelength as $R_{2.90}$. We used the wavelength at 2.90 μm as a proxy for detecting a 3- μm band in Polana's spectra.

The uncertainty in $D_{2.90}$ is:
$$\delta D_{2.90} = D_{2.90} * ((\delta R_1/R_1)\,{}^\wedge 2 + (\delta R_c/R_c)\,{}^\wedge 2\,)\,{}^\wedge 1/2 \qquad (2)$$

Where $R_1 = R_c - R_{2.90}$ \qquad (3)

and $\delta R_1 = (\delta R_c)\,{}^\wedge 2 + (\delta R_{2.90})\,{}^\wedge 2$ \qquad (4)

$\delta R_c$ and $\delta R_{2.90}$ were derived using the uncertainty at each wavelength, computed during the data reduction process. To detect an absorption feature at 2.90 μm (i.e., 3-μm band), $D_{2.90}$ must be greater than $2\delta D_{2.90}$ (2σ).

**Results**

The prism spectrum of Polana exhibits a broad concave feature centered ~1.2 μm with a band depth of ~11%. The spectrum has a slight positive slope toward wavelengths greater than 1.2 μm. We acquired two prism sets of Polana, the first at 12:32 UTC and the second at 12:48 UTC on July 2nd, 2023. Spectra of the two prism sets are similar, showing no compositional heterogeneity in the observed part of Polana. On the other hand, Polana's LXD (1.9– 4.2-μm) spectra do not reveal any pronounced spectral features in the ~2.0-4.0-μm spectral range, suggesting that this asteroid is not hydrated (Figure 2). Using the technique that was described in the method section, the calculated band depth at 2.90 μm is 4.89 ± 7.20 %, ($D_{2.90} < 2*\delta D_{2.90}$), suggesting the lack of a 3-μm band and that the observed surface of Polana is anhydrous (Figure 3).

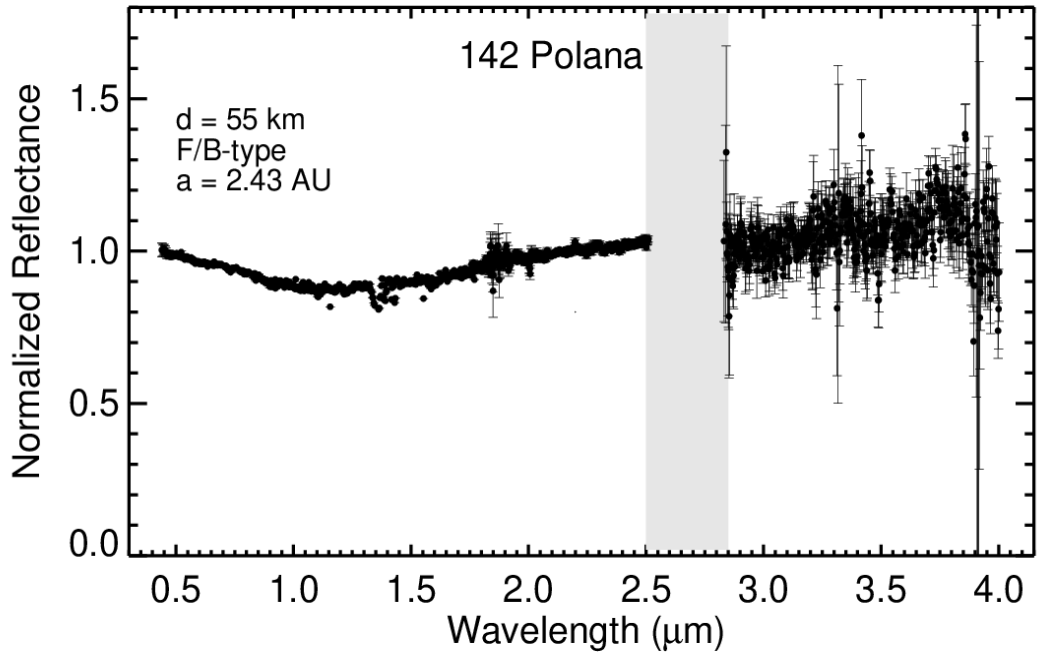

**Figure 2.** The average near-infrared reflectance spectrum of asteroid 142 Polana. The spectrum has been normalized to unity at 2.2 μm. The gray bar marks wavelengths of strong absorption by water vapor in Earth's atmosphere. Uncertainties were computed by Spextool software using the Robust Weighted Mean algorithm with a clipping threshold of 8 (sigma). The value at each pixel is the weighted average of the good pixels, and the propagated variance gives the uncertainty.

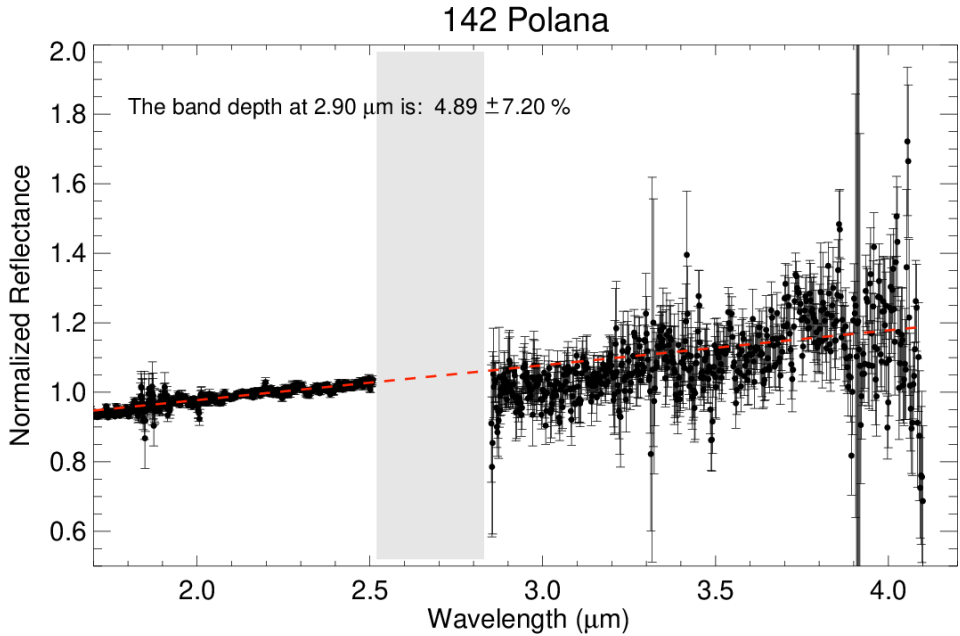

**Figure 3.** The band depth is calculated using a regression line across the K-band of Polana's spectrum. The band depth at 2.90 μm is 4.89 ± 7.20 %, ($D_{2.90} < 2*\delta D_{2.90}$), suggesting the lack of a 3-μm band and that the observed surface of Polana is not hydrated. The gray bar indicates wavelengths of strong absorption by water vapor in Earth's atmosphere. Uncertainties were computed by Spextool software using the Robust Weighted Mean algorithm with a clipping threshold of 8 (sigma). The value at each pixel is the weighted average of the good pixels, and the propagated variance gives the uncertainty.

**Discussion**

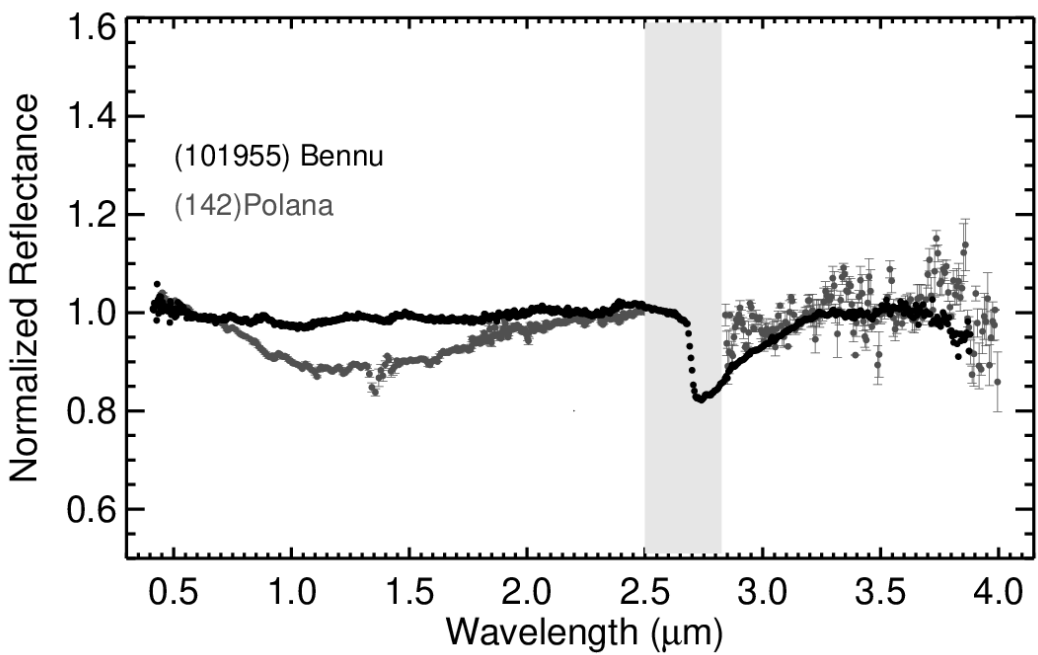

**Figure 4.** Spectra of asteroids Bennu and Polana. The ~1.2-μm band in Polana is more pronounced than in Bennu. Bennu's spectrum is an average spectrum from the equatorial region of the asteroid [28]. Unlike Bennu, Polana does not show a pronounced feature at ~3 μm. Slope-removed spectra are normalized at 2.2 μm.

The prism spectrum of Polana shows a broad feature centered around 1.2 μm (Figure 4), possibly due to amorphous iron-rich silicates, which are abundant in the least-processed CO carbonaceous chondrites that experienced minimal aqueous alteration and thermal metamorphism [29, 30]. Magnetite was also suggested to cause the 1.2-μm feature in B-type asteroids [31]. The 1.2-μm feature in the Polana spectrum is much deeper and more pronounced than Bennu's, suggesting that Polana has more abundant amorphous silicates or magnetite on its surface than Bennu. Polana was previously observed in the 3-μm region by [32], but the relatively low SNR of that spectrum did not allow these authors to confirm the presence of a feature at ~3-μm. In this work, the LXD spectrum of Polana was found to be featureless (does not exhibit a 3-μm feature within 2σ), suggesting that Polana's surface is much less hydrated than Bennu's (Figure 4). Since the duration of LXD observations was approximately 5.5 hours (the rotational period of Polana is about 9.8 hours), the possibility that the unobserved surface of this asteroid is hydrated cannot be excluded. Bennu's spectra were measured by OSIRIS-REx's OVIRS spectrometer, revealing that this primitive asteroid is hydrated with a 3-μm band that has a depth of ~20%, consistent with CM-, CI-, or CR-type carbonaceous chondrites [4]. In addition, the noise in the LXD spectrum is much larger than any feature in 3.4-μm seen on Bennu, whose spectra were found to be consistent with carbonates dominated by calcite and aromatic and aliphatic organics with C-H bonds [33, 34].

If the parent body were uniformly aqueously altered, the lack of pronounced water (OH/$H_2O$), organic materials, and carbonates features on asteroid Polana could be related to the degree of heating produced by impact-generated shocks produced during the New Polana family-forming event, as well as potential heating produced by reaccretion of ejecta that did not reach escape velocity. To lose most of its surface/sub-surface OH-bearing minerals, Polana had to be exposed to impact temperatures higher than ~800 K [35, 36]. Heating generated by disruptive collisions (family-forming events) is substantial, where the temperature of materials influenced by the impact event may increase from 300 K (initial temperature) to 700 K [37]. This result is consistent with numerical hydrocode impact experiments from [38], who suggested that the hydration state of the members of a collisional family can be heterogeneous. The degree of heterogeneity depends on the nature of the impact, the impact energy of the projectile, and how shock heating is distributed within the family.

Several factors can affect the degree of heating and shock metamorphism during family-forming events, including the impactor velocity and size, porosity within the asteroid's parent body, and material ejection efficiency. [39] found that in asteroids, the impactor velocity and size are the main factors responsible for high-grade shock metamorphism in impacts in the Main belt. According to these authors, changing the porosity, accountable for the overall energy absorption within the parent body, from 10% to 30% in their simulations only slightly decreases the shock pressure and temperature.

It is plausible that the heat generated by high-pressure shocks during the New Polana family-forming impact, and the heat produced during ejecta reaccretion, may be sufficient to dehydrate the largest fragment of the New Polana family, asteroid Polana suppress its spectra's 3-$\mu$m absorption feature. The fate of hydration bands on the collisional ejecta escaping during the family-forming event depends on their location near the impact point. Using numerical impact simulations that can track shock heating in the fragments, [39] predicted that only 6% of the debris escaping family forming events (that may go on to produce primitive NEAs) experienced high-impact energy that can degrade their hydration level. Therefore, we expect an aqueously altered parent body to produce ejecta that is also aqueously altered. The nature of the most significant remnant will depend on the heat generated by the impact and ejecta reaccretion.

Another possibility for explaining the discrepancy between Polana and Bennu in the 3-$\mu$m band could be that the parent body had a heterogeneous interior, and the New Polana family-forming event exposed the parent body's sub-surface. The dislodged Bennu fragments from the crustal parent body may contain hydrated silicates, organics, and carbonates, unlike the exposed sub-surface of the parent body, which does not include these materials. This would be consistent with the idea that early heating within the parent body from radiogenic nuclides like [26]Al forces water and other volatiles to move outward toward the exterior of the body, leaving behind relatively dehydrated materials in the deep interior [40].

How about space weathering? Laboratory experiments on carbonaceous chondrites have shown that space weathering can affect spectral characteristics (e.g., [40], [41]). For example, irradiating these carbonaceous chondrites causes their near-infrared spectra to become bluer, brighter, redder, and darker depending on several factors, including their composition, surface grain size, initial albedo, etc. Space weathering can also cause the band depth of mineral absorptions at 0.7 and 3.0 $\mu$m to decrease by 12 % and 50% over a timescale of ~57 Myr and ~7.9 Myr, respectively [42].

The New Polana family is modestly old in the Main Belt, with an estimated dynamical age of $1400 \pm 150$ Myr [3]. This age was determined by modeling the orbital distribution of the observed family members and how they spread over time from the coupled effects of Yarkovsky thermal drift and YORP thermal torques. Once bodies from this family have reached resonances that can take them into the terrestrial planet-crossing region, they tend to have dynamical lifetimes of a few Myr to a few tens of Myr [43].

Based on a comparison between a model crater production function and its largest craters, [44] found that Bennu's surface age ranged between 10 to 65 Myr. Portions of its surface appear considerably younger, however, as determined by the estimated ages of small craters [45]. According to [46], Bennu has been in near-Earth space for 1.75 Myr. However, this value is difficult to determine from Bennu's cratering history because Bennu can be hit by main-belt asteroids while being an NEA, provided its eccentricity is modestly large [47]. Regardless, its surface crater retention age was mainly determined by impacts with main-belt asteroids, whose population is 1000 times larger than the NEA population [3].

Observations of space weathering on Bennu show that the 3-μm band has not been meaningfully affected by space weathering processes over its estimated surface age of 10-65 Myr. Similarly, the fact that many primitive asteroid members in the IMB also show the 3-μm band suggests that space weathering likely does not affect objects that could be as old as the New Polana family itself (e.g., [27]). For this reason, we argue that space weathering cannot explain the discrepancy between Polana and Bennu.

Exogenic materials and breccias are relatively common in meteorites (e.g., Bischoff et al. 2010) and have been seen on Vesta [48, 49]. Exogenous anhydrous materials were also discovered on the rubble pile asteroids Bennu and Ryugu [45, 50]. Much of this contamination probably came from impacts in the asteroid belt over the last 4.5 Gyr. Up to the family-forming event, collisions can mix projectile material into the near-surface of the parent body. At the family-forming event, this exogenic material and the debris from the projectile producing the family itself would be mixed into all the newly-created family members. From there, impacts on the family members could potentially add more foreign materials. This means it is plausible that exogenic materials on Polana, its family members, and Bennu may contribute to its spectra. Accordingly, it cannot be ruled out that exogenic hydrated (in addition to basaltic) materials landed on Bennu, contributing to its hydration level. That said, modeling work suggests that the degree of contamination produced by impacts is usually limited [45]. We expect that the spectral signature of delivered material will not be enough to change a hydrated body to a non-hydrated body, or vice versa, except in some exceptional situations.

Remote sensing observations of the primitive NEAs Ryugu and Bennu suggested that these two asteroids experienced different aqueous alteration histories, as revealed by the characterization of their surface composition using the 3-μm band [4, 51]. The NIRS3 instrument on board Hayabusa2 detected a weak and narrow absorption feature centered around 2.72 μm across the observed Ryugu's surface, attributed to hydroxyl-bearing minerals [51]. Returned samples from Ryugu confirmed that this asteroid has a similar composition to CI-type carbonaceous chondrites [52]. OSIRIS-REx's OVIRS detected a broader and deeper 3-μm band in Bennu compared to NIRS3 observations of Ryugu.

It is possible that Ryugu and Bennu originated from different parent bodies with distinct aqueous alteration and thermal histories, based on their 3-μm hydration features. We find this plausible, with two prominent primitive families in the IMB, New Polana and Eulalia, having reasonable odds of producing Bennu and Ryugu [3]. Although initial findings from Bennu samples have shown that their 3-μm spectra do not match OVIRS' spectra and closely resemble those of Ryugu [54]. [54, 55] identified differences between laboratory spectra, ground-based telescopic, and spacecraft spectra, which may be attributed to factors such as space weathering. We will be able to learn more about the aqueous alteration and thermal histories of Bennu when we analyze its returned samples in detail. With the current ground-based observations of Polana and due to the strong Earth atmospheric absorptions that affect the ~2.7-μm region, it is not feasible to fully assess if this asteroid has a 3-μm band like the narrow and subtle band found in Ryugu. Approved JWST programs to observe Polana (with the NIRSpec and MIRI instruments; program 3760, PI A. Arredondo) and members of the seven primitive IMB (with the NIRSpe instrument; program 6384, PI D. Takir) will allow us to investigate the origin of Bennu and Ryugu further.

## Acknowledgments

We are grateful to the three anonymous reviewers for their valuable feedback, which has enhanced this manuscript. We thank the NASA IRTF staff for their assistance with the Polana asteroid observation. Spextool software is written and maintained by M. Cushing at the University of Toledo, B. Vacca at SOFIA, and A. Boogert at NASA InfraRed Telescope Facility (IRTF), Institute for Astronomy, University of Hawai'i. The University of Hawai'i operates NASA IRTF under contract NNH14CK55B with NASA. D.T. was a visiting astronomer at the Infrared Telescope Facility under contract from the National Aeronautics and Space Administration, operated by the University of Hawaii.

## Data Availability

We declare that the Prism and LXD spectra of asteroid (142) Polana supporting this study's findings are available within the article and its Supplementary data file. In addition, raw data of Polana and corresponding standard stars used for processing will be publicly available in the NASA/IPAC Infrared Science Archive at https://irsa.ipac.caltech.edu/applications/irtf/.